\documentclass[prl,twocolumn,showpacs,preprintnumbers,amsmath,amssymb]{revtex4}

\usepackage[english]{babel}

\usepackage[applemac]{inputenc}
\usepackage[T1]{fontenc} 
\usepackage{hyperref}
\usepackage{graphicx}

\newcommand{\ER}{Erd\H{o}s-R\'enyi }
\newcommand{\mat}[1]{\mathbf{#1}}
\DeclareMathOperator{\nnz}{nnz}

\begin{document}
\title{Unfolding accessibility provides a macroscopic approach to temporal networks}
\author{Hartmut H. K. Lentz}
\affiliation{Institute for Physics, Humboldt-University of Berlin, Newtonstr. 15, 12489 Berlin, Germany}
\affiliation{Institute of Epidemiology, Friedrich-Loeffler-Institute, Seestr. 55, 16868 Wusterhausen, Germany}
\author{Thomas Selhorst}
\affiliation{Institute of Epidemiology, Friedrich-Loeffler-Institute, Seestr. 55, 16868 Wusterhausen, Germany}
\author{Igor M. Sokolov}
\affiliation{Institute for Physics, Humboldt-University of Berlin, Newtonstr. 15, 12489 Berlin, Germany}
\date{\today}

\begin{abstract}
An accessibility graph of a network contains a link, wherever there is a path of arbitrary length between two nodes. We generalize the concept of accessibility to temporal networks.
Building an accessibility graph by consecutively adding paths of growing length (unfolding), we obtain information about the distribution of shortest path durations and characteristic time-scales in temporal networks.
Moreover, we define causal fidelity to measure the goodness of their static representation.
The practicability of our proposed methods is demonstrated for three examples: networks of social contacts, livestock trade and sexual contacts.

\end{abstract}

\maketitle

Many real world systems can be described in terms of networks \cite{Newman_siam}, i.e. sets of nodes, or vertices, and edges connecting node pairs.
Prominent examples are social contacts, flow of people between cities, trade between countries or the world wide web
\cite{Wassermann,Hufnagel,Barabasi,Stanley,Hidalgo}.
In many cases, a static network description is sufficient: an edge is taken present, whenever a connection between the
corresponding nodes was established during the whole data acquisition time. 
Mathematically, static networks are considered as graphs and can be represented by adjacency matrices. 
These contain unit elements in the entries corresponding to the indices of nodes connected by an edge, all other elements being zeroes.
This representation is used in a large number of algebraic methods for network analysis \cite{Newman_book,Aldous_book}.

\begin{figure}[ht]
\begin{center}
\includegraphics{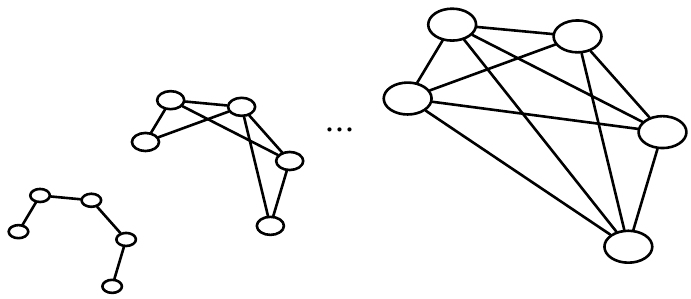}
\caption{Unfolding of an accessibility graph.
Starting with a network (left) we unfold its accessibility graph by adding direct edges for paths of length 2 (center) and so forth.
The fully unfolded accessibility graph is shown on the right.
}\label{fig:unfolding_scheme}
\end{center}
\end{figure}

Every route across the network connecting two vertices along the network's edges is called a path. For a macroscopic view of a network, 
it is often sufficient to know, whether the nodes are connected by a path of a whatever length, i.e. whether one node is accessible starting from another one.
Accessibility can be mapped onto a single mathematical object, the accessibility matrix of the network, 
containing unit entries for indices corresponding to the pairs of the nodes connected by paths of arbitrary length \cite{Newman_book,Aldous_book}.
This matrix is the adjacency matrix of the accessibility graph.
To build this graph, one starts with the network itself and first adds edges between the nodes connected by paths of length 2.
Then one subsequently adds longer paths as edges step-by-step (fig. \ref{fig:unfolding_scheme}), until there are no more new paths to add.
It turns out that valuable insights into the network structure are not only given by the accessibility graph itself, but also by the graphs obtained on intermediate stages of its step-by-step construction.
We call this step-by-step construction the unfolding of the accessibility graph. 

Although the concept of accessibility is plausible in static networks, it is still not clear how to generalize it to networks, where edges are not constant over time.
Many systems have to be represented as temporal networks: their nodes being fixed, but edges, representing connections between them, 
are only present during certain periods of time 
\cite{Vernon,Bajardi:2011,Stehle,Rocha_plosbc,Salathe_highres}, see \cite{Holme2012} for a review.
Temporal networks are not graphs and cannot in general be represented by single matrices.
Therefore, most attention on temporal networks has focussed on data-driven approaches \cite{Rocha_plosbc,Bajardi:2011,Pan}. 
Work concerning macroscopic properties of temporal networks is rare.
This is partly due to the absence of an established formalism to treat these objects \cite{Holme2012,Danon:2011}.
A time-aggregated, static representation of a temporal network may contain paths, which do not follow a causal sequence of connections between nodes, i.e. paths which cannot be taken in a real system.
A static representation may be not sufficient for the description of a temporal network.

The first step for the introduction of a formalism for temporal networks will thus be taking into account the causality of paths.
In the present Letter we introduce accessibility graphs of temporal networks and use them for a description of their macroscopic properties.
In addition, we define the measure of causal fidelity quantifying the goodness of a static representation of a temporal network.
We first briefly review the basic concept of accessibility in static networks and introduce the method of unfolding accessibility for the static case.
Then we discuss generalizations of the concept of accessibility to temporal networks. 
Finally, we analyze three different data sets with respect to their characteristic time scales and compare their temporal and static representations.


Let us consider a static network $G=(V,E)$, where $V$ is the set of vertices and $E$ is the set of edges.
The network is represented by an adjacency matrix $\mat{A}$, with $(\mat{A})_{uv}=1$, if there is an edge between nodes $u$ and $v$ and $(\mat{A})_{uv}=0$, otherwise.
The accessibility graph (transitive closure) of a static network is denoted by $G^*=(V,E^*)$.
It contains an edge $(u,v) \in E^*$, if a path of arbitrary length exists between the nodes $u$ and $v$.
Given an adjacency matrix $\mat{A}$, the elements of the matrix $\mat{A}^n$ are the numbers of paths of length $n$ connecting pairs of nodes of the graph.
Consequently, the cumulative matrix 
$\mat{C}_n = \mat{A}+\mat{A}^2+\cdots +\mat{A}^n$
has a nonzero entry $(\mat{C}_n)_{uv} \neq 0$, if the nodes $u$ and $v$ are connected by a path of whatever length smaller than or equal to $n$.
Summing up all powers of the adjacency matrix, i.e. building a matrix $\mat{C}_{n=N}=\sum _{i=1} ^{N} \mat{A}^i$ (where $N$ is the total number of nodes in the network), 
is related to the transitive closure of the network.

Provided that we are interested only in the number of nodes that can be reached from each initial node within $n$ steps, 
we record whether the elements of $\mat{C}_n$ are zero or not.
Changing every nonzero element of $\mat{C}_n$ to unity yields the accessibility matrix $\mat{P}_n$. 
The matrix $\mat{P}_n$ (with binary entries) can be treated as a Boolean matrix.
Alternatively, one can consider the adjacency matrix $\mat{A}$ as a Boolean matrix from the very beginning.
The product of two Boolean matrices is defined by using normal algebra, but Boolean arithmetics, 
interpreting multiplication as a logical AND ($\wedge $) operation and addition as a logical OR ($\vee $) operation.
The product of two Boolean matrices is 
$(\mat{A}\mat{B})_{ij}= \bigvee _{k=1}^N a_{ik} \wedge b_{kj}$.
In this notation, the accessibility matrix $\mat{P}_n$ of path length $n$ becomes \cite{Warshall:1962}
\begin{equation} \label{eq:staticpathmatrix2}
\mat{P}_n = \bigvee _{i=1} ^n A^i ,
\end{equation}
where the $i$-th power of the boolean matrix $\mat{A}$ is computed using boolean matrix products.
A brief discussion of some properties of the fully exploited accessibility graph $\mat{P}_{n=N}$ is given in the supplementary information \cite{SI}.
The number of edges in $\mat{P}_1$ gives the number of paths of length $1$, and at every step $n \rightarrow n+1$ the network is traversed by paths with one more edge.
Thus, Eq. \eqref{eq:staticpathmatrix2} corresponds to a breadth-first-search (BFS) in the network.
This relation is also used in the Floyd-Warshall algorithm \cite{Warshall:1962,Floyd} for the computation of shortest path lengths.
When the BFS reaches the diameter $D$ of the network, i.e. there are no more shortest paths to traverse, the accessibility matrix saturates, $\mat{P}_N = \mat{P}_D$.

The total number of nonzero elements of the accessibility matrix $\mat{P}_n$ is associated with the distribution of shortest path lengths.
We define the density $\rho(\mat{A})$ of an $N\times N$ matrix $\mat{A}$ with $N^2$ entries and $\nnz (\mat{A})$ nonzero elements as
\begin{equation}
\rho(\mat{A})= \frac{\nnz (\mat{A})}{N^2}.
\end{equation}
When $\mat{A}$ is an adjacency matrix, $\rho (\mat{A})$ defines the edge density of the network, giving the probability to find an edge between two nodes chosen at random. 
In analogy, the probability to find a path of length $l \leq n$ between two randomly chosen nodes in a connected network is given by
\begin{equation}\label{eq:cdf}
F_n = P(l\leq n) \equiv \rho (\mat{P}_n).
\end{equation}
Consequently, the \emph{path density} $\rho (\mat{P}_n)$ gives the cumulative distribution of shortest path lengths in the network.
The corresponding probability to find the shortest path of length $n$ is given by $f_n=(F_n - F_{n-1})$, with $F_0=0$.
The distribution saturates, when $n$ reaches the diameter $D$ of the network.
For a single-component network $F_D = \rho(\mat{P}_D) = 1$.
For networks comprised of disjoint components $\rho(\mat{P}_D) <1$, since their accessibility matrices have less than $N^2$ nonzero entries \cite{SI}.
In this case we consider $F_n$ as an ``improper'' probability distribution, which is not normalized to unity, and
define its median as the value of $n$, where $F_n$ approaches one half of its saturation value $F_D$.

Temporal networks allow edges to vary over time, and are
represented by a sequence of adjacency matrices $\mathcal{A}=\{\mat{A}_1, \mat{A}_2,\dots , \mat{A}_T \}$,
where each matrix $\mat{A}_t$ is a snapshot of the system at time $t$.
The increment corresponds to the temporal resolution and $T$ is the maximal time given in the data.
The accessibility graph $\mathcal{G} ^*=(V,E^*)$ of a temporal network has to allow for 
only causal paths (time respecting paths), in which the temporal order of edges has to be taken into account: it 
contains an edge $(u,v) \in E^*$, if and only if there is a causal path between the nodes $u$ and $v$.
A time-aggregated, static representation follows by summing up all (Boolean) matrices in the sequence, $\mat{A}=\vee_{t=1}^T \mat{A}_t$.

As we have seen for the static case, the product of two identical adjacency matrices gives information about the number of paths of length two.
If we multiply different matrices, i.e. two matrices $\mat{A}_{1}$ and $\mat{A}_{2}$ in the sequence $\mathcal{A}$ we obtain non-zero entries, wherever nodes receive 
edges at time $1$ and cast forth edges at time $2$.
It is important to emphasize that the entries $(\mat{A}_{1} \mat{A}_{2} ) _{ij} $ vanish, if an edge received by node $i$ at time $1$ is not cast forth to node $j$ at exactly time $2$.
Using products of different adjacency matrices in the sequence $\mathcal{A}$, we can derive a dynamic accessibility matrix exactly as it was done in the static case.
This straight-forward approach, however, does not give useful results, because it does not include the possibility of node waiting times (see supplementary information \cite{SI}).

To overcome this drawback, we allow for temporal shortcuts in the accessibility matrix by taking into account products of adjacency matrices corresponding to remote time steps.
As an example, if many nodes in the system receive links at time $5$ and cast forth links at time $100$, the product $\mat{A}_5 \mat{A}_{100}$ should make a large contribution to the accessibility matrix.
To introduce shortcuts, we define the accessibility matrix of a temporal network by adding an identity matrix $\mat{1}$ to each matrix $\mat{A}_i$ in the sequence.
This yields all ordered products of snapshot matrices in time: $ \mathcal{C}_n=\prod _{i=1} ^n (\mat{1} + \mat{A}_i )  $.
%
The addition of an identity matrix can be interpreted as the ability of each node to keep edge information over time, i.e. it introduces memory into the system. 

Since we are not interested in the actual number of paths between two nodes, we can either set any nonzero element of $\mathcal{C}_n$ to unity or, 
equivalently, use the Boolean formulation
\begin{align}
\mathcal{P}_n=&\bigwedge _{i=1} ^n (\mat{1} \vee \mat{A}_i ) \label{eq:temporalpathgraph_bool} \\
= & \mat{1}\vee \mat{A}_1\vee \mat{A}_2\vee \mat{A}_3\vee \cdots \vee \mat{A}_n \vee \nonumber \\
& \vee \mat{A}_1\mat{A}_2\vee \mat{A}_2\mat{A}_3\vee  \mat{A}_1\mat{A}_3\vee  \mat{A}_1\mat{A}_2\mat{A}_3 \vee \cdots . \nonumber
\end{align}
Linear terms in \eqref{eq:temporalpathgraph_bool} correspond to the aggregated network.
Higher order products (always respecting a causal sequence of indices) represent all possible direct connections or temporal shortcuts.
It should be noted that, due to the causality of the paths, the temporal accessibility graph of an undirected network is in general directed.
The unfolding of the accessibility matrix given in \eqref{eq:temporalpathgraph_bool} can now be used to compute shortest-path distributions $F_n$ for temporal networks 
in the same manner as for the static case.
Note that the index $n$ has now the meaning of the duration of the path (total number of time steps necessary to traverse the system, regarding waiting times).
It defines the time (in units determined by the temporal resolution of the used data) which, say, the passengers spent in motion. 
Thus, a temporal network may still show a small-world property as long as its time-aggregated representation is a small-world network.
However, if the the links between nodes are activated only during distant periods in time, 
the spread over the network would be slow: even a small world can be a ``slow world''. 

We have to point out that the identity matrix on the right-hand side of \eqref{eq:temporalpathgraph_bool} is an artifact of the introduced memory term.
It does not make any difference in undirected networks, whose accessibility matrices always exhibit non-vanishing diagonal entries for $n\geq 2$ \cite{SI}, but may make a difference in directed networks.
However, since the number of diagonal elements is small compared 
to the total number of elements of a large matrix, this deviation is negligible in large networks.  

Temporal networks are often approximated by their time-aggregated counterparts.
However, some paths in a static network representation do not follow a chronological order and are never taken. 
In order to quantify the quality of a static representation, we define the \emph{causal fidelity} $c$ as the fraction of the number of paths in a static network which can be also taken in a temporal one:
\begin{equation}\label{eq:causality}
c =\frac{\rho (\mathcal{P}_T)}{\rho (\mat{P}_T) } .
\end{equation}
The values of $c$ lay in the interval $0\leq c \leq 1$, where $c=1$ means that the path density of a static representation does not differ from the temporal system.
Low values of $c$ indicate that the majority of paths in the static approximation do not follow a causal sequence of edges and thus are not present in the temporal one.

As examples we consider several real-world networks showing different time scales ranging from three days \cite{isella2011} 
to six years \cite{Rocha_pnas}, see Table ~\ref{tab:data}.
\begin{table}[htbp]
\centering
\caption{Temporal network data sets.}
\begin{tabular*}{\hsize}{@{\extracolsep{\fill}}lrrrrl}
Data set & size  & snapshots & $t$-res. & type & source \cr
\hline
Conference & 113  & 10618 & 20 s & undirected & \cite{isella2011} \cr
Pig trade & 89745 & 465 & 1 d & directed &\cite{Euro-Lex} \footnote{www.hi-tier.de. Data access restricted to competent authorities in Germany.} \cr
Sexual contacts & 16730 & 2232 & 1 d & undirected & \cite{Rocha_pnas} \cr
\hline
\end{tabular*}
\label{tab:data}
\end{table}

\begin{figure}[htbp]
\begin{center}
\includegraphics{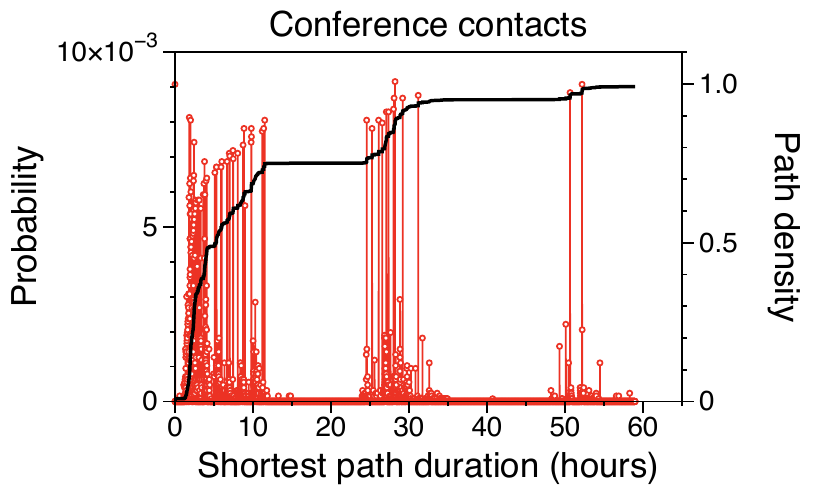}
\caption{(Color online) Density of the accessibility graph (black line) and distribution of shortest path durations (red) for a network of face-to-face contacts at a conference.
}
\label{fig:cumu_histo_113}
\end{center}
\end{figure}

\begin{figure}[h!]
\begin{center}
\includegraphics{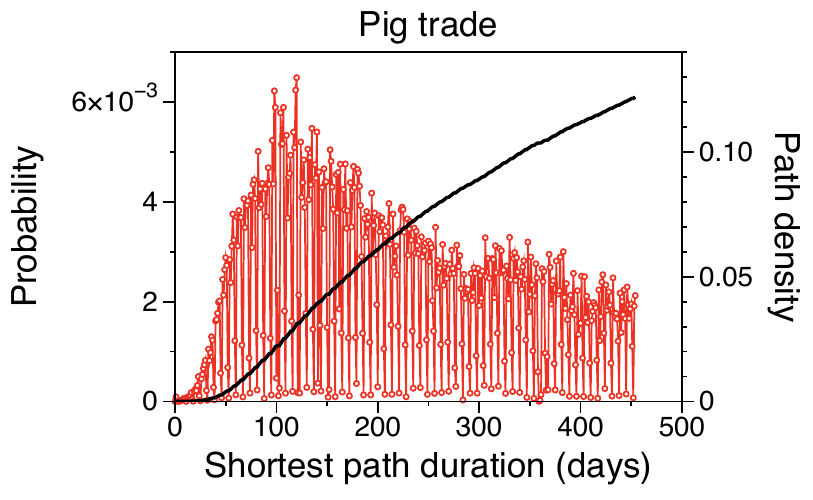}
\caption{(Color online) Density of the accessibility graph (black line) and distribution of shortest path durations (red) for a network of pig trade in Germany.
}
\label{fig:cumu_histo_pig}
\end{center}
\end{figure}

\begin{figure}[h!]
\begin{center}
\includegraphics{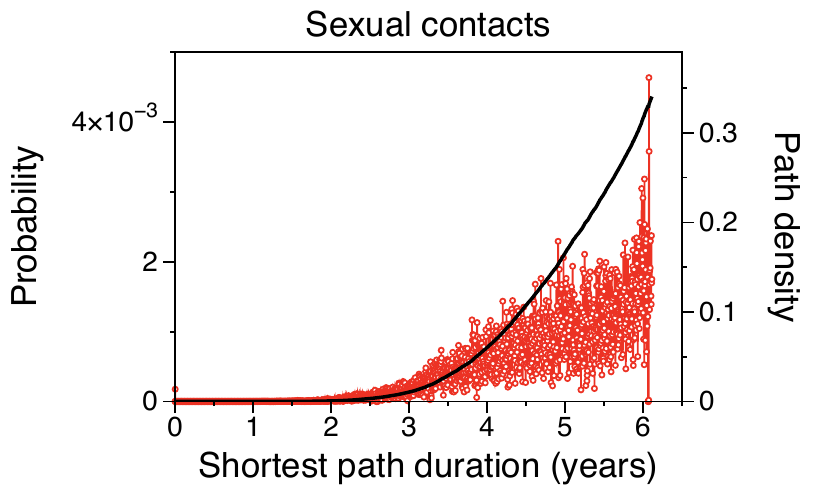}
\caption{(Color online) Density of the accessibility graph (black line) and distribution of shortest path durations (red) for a network of sexual contacts.
}
\label{fig:cumu_histo_sexual}
\end{center}
\end{figure}

We unfold the accessibility graphs of those networks and obtain path densities and distributions of shortest path durations.
Hereby, we treat the path density as an (improper) cumulative distribution.
The results are shown in figures \ref{fig:cumu_histo_113}, \ref{fig:cumu_histo_pig} and \ref{fig:cumu_histo_sexual}.

Figure \ref{fig:cumu_histo_113} shows results for a network of conference attendees.
Every face-to-face contact between two persons is recorded as an undirected edge.
The data set covers a time span of 3 days, separated by time spans of no activity (nights).
The density of the accessibility graph in figure \ref{fig:cumu_histo_113} shows clear and fast saturation at a value of $\rho = 0.9917$, 
indicating that after only 3 days a large part of the network is causally connected.
More than 70\% of the possible paths are traversed within only one day (black line).
The median of the distribution is reached in less than 6 hours.
An aggregated network representation gives 
a good approximation of the system, which is also reflected in the high causal fidelity of $c\approx 0.99$ of its static representation.

Figure \ref{fig:cumu_histo_pig} shows the results for the network of pig trade in Germany over a time span of 455 days.
Every node in the network is an agricultural holding and an edge is present, when a holding sells livestock animals to another one.
In contrast to the two other data sets, the pig trade network is directed.
This is reflected in relatively low values of the path density.
The shortest path lengths show a broad distribution (red).
Although the temporal diameter of the pig trade network is by far larger than the observation period of 455 days, the distribution shows a global maximum (i.e. possesses a mode) at around 120 days.
This means that the typical spreading time-scales are of the order of 100 days. 
The comb-like structure of the graph reflects the time-inhomogeneity of trade, which is not so frequent on saturdays and almost absent on sundays.
The static representation of the network captures its features sufficiently well ($c=0.72$) and can be in principle used in epidemiological modeling.
The same behavior can be expected for pig trade networks of other countries, because it is determined by the typical lifetime of a livestock pig.

Figure \ref{fig:cumu_histo_sexual} shows the distribution of shortest path durations for a network of sexual contacts.
The network reflects physical contacts between prostitutes and their customers.
Each contact is recorded as an undirected link.
Here, no saturation of the path density is reached within the study period, and 
the distribution of shortest path durations does not show a clear maximum.
The figure demonstrates that the majority of shortest paths takes more than 2 years to traverse.
Even if time-scales cannot be resolved from the data, the observation is useful 
to define a minimum duration of future field observations and may be of use for the implementation of vaccination strategies.
The figure shows that it takes at least 3 years to obtain a considerable density of the accessibility graph.
The sexual contact network is only poorly represented by its aggregated counterpart, as
reflected by low causal fidelity of this representation ($c=0.38$).
A static network would significantly overestimate the number of transmission paths, and the system has to be analyzed 
from a temporal network perspective \cite{Rocha_pnas}.

To conclude, we introduced temporal accessibility graphs as a formalism to analyze temporal network data in terms of causal paths.
The method is the temporal generalization of the transitive closure of a static network.
It gives information about the distribution of time-scales in the system and also provides an instrument for quantifying the quality of their simpler, static representation.
The examples of real networks considered demonstrate the practicability and usefulness of this approach. 

This research was funded by the Federal Ministry of Education and Research, grant 13N9520.

\newpage
\section{Supplementary Material}
\section{Properties of static accessibility graphs}
We briefly report some properties of the fully exploited accessibility graph 
\begin{equation}\label{eq:staticpathmatrix}
\mat{P}_{n=N} = \bigvee _{i=1} ^N A^i,
\end{equation}
where $N$ is the total number of nodes in the network.
If the network is connected, $\mathbf{P}_N$ is a matrix with all unit entries, i.e. it has $ \sum _{ij} (\mathbf{P}_N)_{ij}=N^2$ nonzero elements.
In networks consisting of disjoint components, $\mathbf{P}_N$ can be transformed to a block-diagonal form, with blocks filled with unit entries.
Due to the block structure, the matrix has less than $N^2$ nonzero entries in this case.
In directed networks, the diagonal entries of $\mat{P}_n$ can be $0$ or $1$ for any $n$.
The diagonal entries of $\mat{P}_n$ in undirected connected networks are unities for all $n\geq 2$, because there is always a path of length 2 from any node in the network back to itself.

The distance between two nodes is the length of the shortest path connecting them.
The diameter $D$ of the network is the maximum distance between any two nodes in a network.
Increasing $n$ in equation \eqref{eq:staticpathmatrix} successively adds longer shortest paths into $\mat{P}_n$.
When $n$ approaches the diameter of the network, no new nodes can be reached by considering longer shortest paths.
Thus, the diameter $D$ of a network is the minimal number so that $\mathbf{P}_{D+1} = \mathbf{P}_D=\mathbf{P}_N$.

\section{Example: Unfolding accessibility of a directed static \ER graph}
\begin{figure}[h!]
\begin{center}
\centerline{\includegraphics{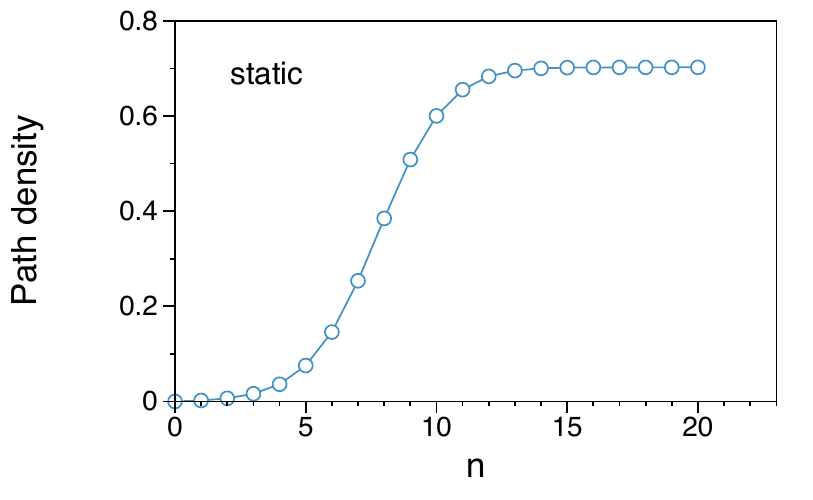}}
\caption{Density of the accessibility graph of a directed random network vs. $n$.
The path density saturates at $n=18$ (diameter of the network).
Network: 1000 nodes, 2000 directed edges.}\label{fig:ER_cumu}
\end{center}
\end{figure}

As an example, we plot the path-density $\rho (\mat{P}_n)$ for a directed \ER graph $G_e=(V,E)$ with 1000 nodes and 2000 edges in figure \ref{fig:ER_cumu}.
The curve saturates for $n>18$ indicating that $18$ is the diameter of the network.
The saturation value of $\rho _\text{sat}(\mat{P}_n) \approx 0.7$ is smaller than unity, since the network $G_e$ is not strongly connected.
The median of path lengths is located at $n=8$, where the density approaches $\rho (\mat{P}_n) \approx 0.35$.
The corresponding shortest path length distribution $f_n$ shows that the shortest path 
length is on average much smaller than the number of nodes in the network (figure \ref{fig:ER_cumu}) indicating the
small-world nature of the network.

%

\begin{figure}[h!]
\begin{center}
\includegraphics{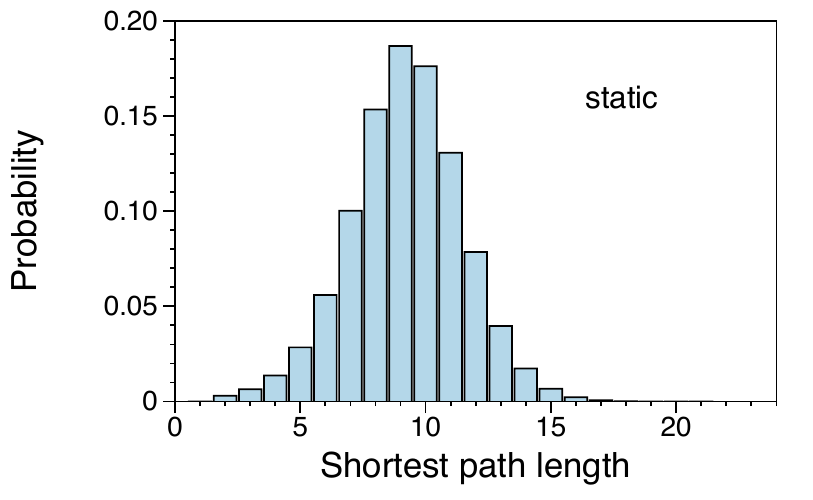}
\caption{Shortest path length distribution as it follows from fig. \ref{fig:ER_cumu}.
Mean value = 8.18.
}
\label{fig:ER_histo}
\end{center}
\end{figure}

\section{Straight-forward temporal generalization of static accessibility}
A straight-forward way of generalizing the accessibility matrix to a system of the form
\begin{equation}\label{eq:seq}
\mathcal{A}=\mat{A}_1, \mat{A}_2,\dots , \mat{A}_T
\end{equation}
could consist in replacing the powers of the same matrix $\mat{A}$ by products of different matrices $\mat{A}_t$:
\begin{equation}\label{eq:wrong}
\mathcal{C}_n =\sum _{i=1} ^n \prod _{j=1} ^i \mat{A}_j = \mat{A}_1+\mat{A}_1\mat{A}_2+ \mat{A}_1\mat{A}_2\mat{A}_3 + \cdots 
\end{equation}
and
\begin{equation}\label{eq:wrong_bool}
\mathcal{P}_n =\bigvee _{i=1} ^n \bigwedge _{j=1} ^i \mat{A}_j = \mat{A}_1\vee \mat{A}_1\mat{A}_2\vee \mat{A}_1\mat{A}_2\mat{A}_3 \vee \cdots .
\end{equation}
Every higher order product $\mat{A}_k \cdots  \mat{A}_m$ in \eqref{eq:wrong_bool} contributes to $\mathcal{P}_n$, if and only if there is a chronological 
sequence of edges placed at exactly the time steps $k \dots m$.

In many systems, however, there are waiting times between two link activations.
As an example, many nodes of the network could receive links at time $1$ and be idle until they cast forth links at time $3$.
This factor plays a role in transportation or information networks, where passengers or information are transported to a node $v$ and do not disappear, but wait for one time step, and are then transported further at time step 3.
In the mathematical context, this means that the product $\mat{A}_1 \mat{A}_2$ could yield zeros for a row/column $v$, but the product $\mat{A}_1 \mat{A}_3$ would not.
These waiting times cannot be captured by the approach suggested by our simple ansatz. This drawback is reflected in the fact that the products of higher order vanish in sparse networks for purely statistical reasons, i.e.
\begin{equation}\label{eq:limit_zero}
\lim _{n\rightarrow \infty} \prod _{i=1} ^n \mat{A}_i =\mat{0}.
\end{equation}
A dilution of the temporal network occurs, when the intrinsic node time scale (node waiting time) is much larger, than the edge dynamic (temporal resolution).
An example was reported in \cite{Bajardi:2011}, where the maximum path-length for immediately consecutive edges was found to be 8 (days) for a network of cattle trade.
However, the intrinsic timescale for cattle trade is in the order of years.

\bibliographystyle{elsarticle-num-names}
\bibliography{lentz.bib}

\end{document}